# KU-ISPL SPEAKER RECOGNITION SYSTEMS UNDER LANGUAGE MISMATCH CONDITION FOR NIST 2016 SPEAKER RECOGNITION EVALUATION

*Suwon Shon, Hanseok Ko*

School of Electrical Engineering, Korea University, South Korea
swshon@ispl.korea.ac.kr, hsko@korea.ac.kr

## ABSTRACT

Korea University – Intelligent Signal Processing Lab. (KU-ISPL) developed speaker recognition system for SRE16 fixed training condition. Data for evaluation trials are collected from outside North America, spoken in Tagalog and Cantonese while training data only is spoken English. Thus, main issue for SRE16 is compensating the discrepancy between different languages. As development dataset which is spoken in Cebuano and Mandarin, we could prepare the evaluation trials through preliminary experiments to compensate the language mismatched condition. Our team developed 4 different approaches to extract i-vectors and applied state-of-the-art techniques as backend. To compensate language mismatch, we investigated and endeavored unique method such as unsupervised language clustering, inter language variability compensation and gender/language dependent score normalization.

## 1. INTRODUCTION

This document is description of the Korea University – Intelligent Signal Processing Laboratory (KU-ISPL) speaker recognition system for NIST 2016 speaker recognition evaluation (SRE16).

Under i-vector framework, new approaches are introduced using Bottleneck Feature (BNF) and Deep Neural Network (DNN) which were validated successfully its performance improvement on ASR. In this study, we developed the state-of-the-art i-vector systems for validating the performances on language mismatch condition using SRE16 dataset. Based on the prior studies about domain adaptation and compensation, Inter Dataset Variability Compensation (IDVC) and unsupervised domain adaptation using interpolated PLDA are also applied.

After studying about prior works, we proposed additional techniques for compensating the language mismatch condition to obtain robust performance on SRE 16 dataset. For official evaluation, we submitted total 3 systems including 1 primary system and 2 contrastive systems in fixed training data condition. We carefully followed the SRE16 rules and requirements during training and test processes.

In the following, we introduce a dataset of SRE 16 at section 2. At Section 3 and 4, system components for development of state-of-the-art i-vector extraction are described.

## 2. DATASET PREPARATION FOR FIXED TRAINING CONDITION

For fixed training condition, we use Fisher English, SRE 04~10 and SWB-2 (phase1~3, cellular 1~2) dataset for training set. Language of all dataset in training set is English. The dataset for SRE 16 evaluation trials are collected from speakers who located outside North America and spoke Tagalog and Cantonese (referred as *major* language). Before evaluation dataset is available, development dataset which mirrors the evaluation conditions to prepare the language mismatch condition on evaluation set. The development dataset is collected from speaker who located outside North America and spoke Cebuano and Mandarin (referred as *minor* language). Additionally, unlabeled minor and major language dataset is also given to participants for development set. The development set are free to use for any purpose and detailed statistics about evaluation and development dataset are shown in table 1.

**Table 1**. Statistics of development and evaluation dataset. * means information from the SRE16 plan documents.

| Data set | Category | Language | Labels (metadata) | Numbers of | | |
|---|---|---|---|---|---|---|
| | | | | Utt. | Spk. | Calls |
| Dev. | Enrollment | Minor | Available | 120 | 20 | 60 |
| | Test | Minor | Available | 1207 | 20 | 140 |
| | Unlabeled | Minor | X | 200 | 20* | 200* |
| | Unlabeled | Major | X | 2272 | X | X |
| Eval | Enrollment | Major | X | 1202 | 802 | 602 |
| | Test | Major | X | 9294 | X | 1408 |

## 3. SYSTEM COMPONENT DESCRIPTION

### 3.1. Acoustic features

For training speaker recognition system on this paper, Mel-Frequency Cepstral Coefficients (MFCC) is used to generate 60 dimensional acoustic features. It is consist of 20 cepstral coefficients including log-energy C0, then, it is appended with its delta and acceleration. For training DNN based acoustic model that is inspired by Automatic Speech Recognition (ASR) area, different configuration was adopt to generate 40 ceptral coefficient without energy component for high resolution of acoustic features (ASR-MFCC). For feature normalization, Cepstral Mean Normalization is applied with 3 seconds-length sliding window.

After extracting acoustic features, Voice Activity algorithm was adopted to remove silence and low energy segments on the speech dataset. Simple energy based VAD was used with log-mean scaled threshold. Using log-energy (C0) component of MFCC, the mean log-energy of each segment can be calculated and it is scale to half value and then plus by 6. That is the final scaled threshold for VAD. We do not apply VAD algorithm when we training DNN acoustic model.

### 3.2. I-vector extraction

For performance comparison of SRE 16 trials, four different approaches to extract i-vectors are developed.

#### 3.2.1. GMM-UBM (GU)

For General i-vector extraction approach [1] by modifying Kaldi's recipe (sre08/v1). For training GMM-UBM and total variability matrix, SRE(04~10, part of 12) and switchboard dataset (p2 1~3, cellular 1~2) were used.

#### 3.2.2. DNN-UBM (DU)

Based on Kaldi's recipe (sre10/v2), Fisher English was used for training Time Delay Neural Network with ASR-MFCC feature. After training TDNN, the DNN-UBM is estimated on DNN-MFCC feature which is high resolution version of MFCC. SRE (04~10, part of 12) and Switchboard Dataset were used for training DNN-UBM and total variability matrix [2].

#### 3.2.3. Supervised GMM-UBM (SU)

Based on Kaldi's recipe (sre10/v2), Supervised GMM-UBM[2] was trained using posterior of TDNN network. Same dataset was used as GMM-UBM system for training Supervised GMM-UBM and total variability matrix

#### 3.2.4. Bottleneck Feature based GMM-UBM (BU)

BNF features were extracted using DNN which containing bottleneck layer [3], [4]. DNN layer structure was set to 1500-1500-80-1500 with total 4 layer and MFCC feature of all dataset was converted to BNF feature (80 dim). After extracting BNF feature, it follows general GMM-UBM based i-vector extraction approaches such as GMM-UBM system at Sec. 3.2.1 and same dataset was used for GMM-UBM total variability matrix.

### 3.3. Backend procedures

#### 3.3.1. Inter Dataset Variability Compensation (IDVC)[5]

SRE and Switchboard (SWB) Dataset sub-corpora label and gender label are used for obtaining the average i-vectors of each dataset by gender. SRE can be divide in to 5 sub-corpora (SRE-04, 05, 06, 08, 10) and SWB can be divide in to 5 sub-corpora (switchboard-2 phase 1,2,3 and cellular part 1, 2). Finally, 600 dimensional i-vectors projected to 580 dimension by removing dataset dependent dimension.

#### 3.3.2. Whitening Transform and Length Normalization using unlabeled dataset (WTLN)[6]

Whitening transformation and length normalization are simple and powerful techniques to improve performance of speaker recognition system by compensating the mismatch between enrollment and test i-vector length. It became a mandatory process of i-vector based speaker recognition system back-end and, moreover, recent study validated its effectiveness on domain adaptation by calculating whitening transform matrix using the in-domain dataset. We use use both unlabeled minor and major dataset for whitening and length normalization.

#### 3.3.3. Interpolated PLDA (SRE04-08) + PLDA (speaker clustering using AHC) (IPLDA) [7]

Agglomerative Hierarchical Clustering approach for unlabeled in-domain datasets to estimate the PLDA model was introduced by Garcia-Romero. Using the clustered speaker information, in-domain Within-speaker Covariance (WC) and Across-speaker Covariance (AC) of PLDA model are interpolated from out-of-domain WC and AC. We applied this approach on the unlabeled dataset of minor and major language. By experiments, 30 and 450 clusters were used for speaker clustering of unlabeled minor and major dataset. The 450 clusters (speaker) information of unlabeled major dataset could be used for calibration as Sec.4.5.

#### 3.3.4. S-norm [8]

Symmetric normalization(S-norm) is adopted for score normalization. Basically, unlabeled major dataset was used as imposter utterances for both development and evaluation trials.

## 4. STUDIES FOR COMPENSATING LANGUAGE MISMATCH

### 4.1. Gender Classification and unsupervised Language Classification of minor/major unlabeled dataset (GCLC)

Gender classification could be done by comparing cosine similarity between gender i-vector and input i-vector which we want to classify the gender. Gender i-vector obtained by averaging the i-vectors of training set by gender.

Language classification can be done by unsupervised clustering algorithm such as AHC or k-means clustering. Since k-means clustering performance is greatly depend on initial point as figure 1, AHC is frequently used on i-vector feature space.

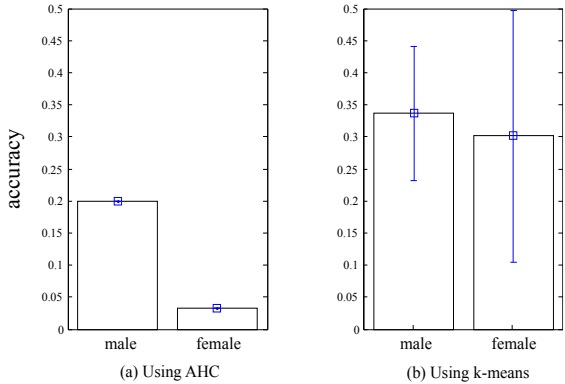

Figure 1. Unsupervised language classification using enrollment dataset of minor language

For high accuracy and reliability of clustering algorithm, we proposed 2-step approach by running k-means algorithm twice on different i-vector representation space. If the k-means algorithm secure good initial clusters that represent the mean of i-vectors from each language, we can get a high language classification performance while reducing the risk of misclassification by random initial cluster.

We check this on the minor enrollment and test dataset in development set which have language labels, so that they allowed us to investigate the performance of language classification. We use 2-step approach for language clustering as below schemes.

    1. Initializing on IDVC subspace (alike PCA)
    2. K-means
           Or
    1. Initialing using AHC
    2. K-means

Both approaches shows same result on minor enrollment and test dataset and it showed 100% language clustering accuracy on minor enrollment dataset and 99.8% on minor test dataset. Using this 2-step approach, we classify the unlabeled minor and major dataset to discover the language label. These valuable information were used very effectively for language mismatch compensation and score normalization in next sections.

### 4.2. Inter Language Variability Compensation for gender and minor/major language (ILVC)

If the system has gender and language information, inter language variability factor can be removed by the same scheme of IDVC. By the high performance of GCLC approach that we proposed in section 4.1, we could have valuable gender and language labels of minor/major unlabeled dataset. Finally, the i-vector subspace removal can be done by the mean i-vector of 10 sub-category according to language (English, Cebuano, Mandarin, Tagalog and Cantonese) and gender.

### 4.3. Simplified Autoencoder based Domain adaptation (SADA)

The Autoencoder based Domain adaptation (AEDA) was proposed recently and its paper is in peer-reviewing process for publication. On this study, we simplify the AEDA to more simple Autoencoder structure and proposed a Simplified Autoencoder based Domain Adaptation (SADA), but it still performs almost same as AEDA. More details about SADA can be found in next studies.

### 4.4. Gender and Language dependent score normalization (GL-Norm)

We have gender and language information of unlabeled dataset by GCLC approach in section 4.1. So we divide S-norm parameter into 4 sub-categories by gender and language. Gender and language of input i-vector are also classified by GCLC approach and use appropriate parameters to get gender and language specific score normalization.

### 4.5. Calibration and fusion

For calibration and fusion, simple linear calibration and linear fusion were done by Bosaris toolkit [9]. For calibration, speaker clustering information of unlabeled major dataset was used (see Sec. 3.3.3) to obtain target and non-target score distribution of evaluation experiments. The mean of speaker cluster represent speaker i-vector and they can be scored with unlabeled major i-vectors. As we have already speaker label from speaker clustering, we can obtain target score and non-target score distribution and they could be used for score calibration on evaluation trials. Calibration was done on both before and after score normalization.

## 5. SUBSYSTEMS FOR MINOR LANUGUAGES

By applying the components described in section 3 and 4, we evaluate the development trials in terms of EER, minimum $C_{primary}$, and actual $C_{primary}$. From the experiments on development trials, we confirm that the propose ILVC and GL-norm approach works better than prior works in language mismatch conditions.

Table 2. Performance evaluation minor language speaker recognition system on development trials

| System Name (i-vector and applied techniques) | S-norm | | | GL-norm | | |
|---|---|---|---|---|---|---|
| | EER | $minC_{primary}$ | $actC_{primary}$ | EER | $minC_{primary}$ | $actC_{primary}$ |
| GU-IDVC-WTLN-IPLDA | **18.3927** | 0.7017 | 0.7153 | 18.3720 | **0.7110** | **0.7239** |
| DU-IDVC-WTLN-IPLDA | 18.8587 | **0.6935** | **0.7057** | **18.3513** | 0.7114 | 0.7314 |
| SU-IDVC-WTLN-IPLDA | 19.9720 | 0.7109 | 0.7281 | 19.5112 | 0.7140 | 0.7336 |
| BU-IDVC-WTLN-IPLDA | 21.0128 | 0.7404 | 0.7718 | 20.5727 | 0.7418 | 0.7804 |
| DU-SADA-WTLN-IPLDA | 19.7494 | 0.7272 | 0.7408 | 19.3662 | 0.7254 | 0.7502 |
| Fusion of 5 sub-systems | 16.7357 | 0.6253 | 0.6347 | 16.4095 | 0.6345 | 0.6396 |
| GU-IDVC-ILVC-WTLN-IPLDA | **16.4043** | 0.6849 | 0.7024 | **16.4872** | 0.6790 | 0.6881 |
| DU-IDVC-ILVC-WTLN-IPLDA | 17.0568 | **0.6454** | **0.6702** | 16.9221 | **0.6346** | **0.6515** |
| SU-IDVC-ILVC-WTLN-IPLDA | 17.6471 | 0.7075 | 0.7113 | 17.4814 | 0.6837 | 0.6930 |
| BU-IDVC-ILVC-WTLN-IPLDA | 18.3927 | 0.7197 | 0.7431 | 18.2425 | 0.7074 | 0.7336 |
| DU-SADA-ILVC-WTLN-IPLDA | 18.0768 | 0.7040 | 0.7112 | 17.8749 | 0.6807 | 0.7053 |
| Fusion of 5 sub-systems | 13.8567 | 0.5800 | 0.5839 | 13.53 | 0.5651 | 0.5742 |

## 6. SUBSYSTEMS AND ITS FUSION FOR MAJOR LANGUAGE

We try to use development dataset as much as possible because the development set is mirror the evaluation, so it would contain more valuable information than training dataset which language is English.

Each subsystems has been applied most competitive techniques from the studies in previous sections for best result. Contrary to estimating the gender and language labels of unlabeled dataset for ILVC at Sec.5 and table 2, we use enrollment and test dataset of minor language and its labels for ILVC. According this method, the performance of subsystems improved dramatically as table 3. Main reason is that we did not process each trials independently and use development enrollment and test dataset information on development trials. We expect that the performance of evaluation trials would not be improved dramatically like development trials, however, still convince that it would influence beneficial effects on systems for major languages. Submitted system has little difference on score normalization and usage of minor language labels.

The primary system is fusion of 5 subsystems of top 5 rows in table 3. We used entire minor dataset (enrollment, test and unlabeled) and unlabeled major dataset for GL-norm on both development and evaluation experiments.

For Contrastive 1 system, only unlabeled major dataset is used for GL-norm on both development and evaluation experiment. Ground truth label was not used for ILVC, so the result is same with the last row table. 2 at GL-norm tab.

Table 3. Performance evaluation major language speaker recognition system on development trials. * means algorithm conducted using true label of enrollment dataset in minor language

| | Unequalized | | | Equalized | | |
|---|---|---|---|---|---|---|
| | EER | $minC_{primary}$ | $actC_{primary}$ | EER | $minC_{primary}$ | $actC_{primary}$ |
| GU-IDVC-ILVC* -WTLN-IPLDA-GLnorm | **12.5621** | 0.6548 | 0.6713 | 12.51 | 0.6416 | 0.6554 |
| DU-IDVC-ILVC* -WTLN-IPLDA-GLnorm | 13.5408 | **0.6097** | **0.6174** | **11.12** | **0.5785** | **0.5915** |
| SU-IDVC-ILVC* -WTLN-IPLDA-GLnorm | 13.9654 | 0.6775 | 0.7023 | 11.78 | 0.6678 | 0.6935 |
| BU-IDVC-ILVC* -WTLN-IPLDA-GLnorm | 14.6127 | 0.7269 | 0.7404 | 12.53 | 0.7077 | 0.7251 |
| DU-SADA-ILVC* -WTLN-IPLDA-GLnorm | 14.1466 | 0.6316 | 0.6450 | 11.53 | 0.6036 | 0.6159 |
| Primary | 9.55 | **0.4913** | 0.5066 | 7.70 | 0.4740 | 0.4882 |
| Contrastive 1 | 13.53 | 0.5651 | 0.5742 | 12.25 | 0.5940 | 0.6065 |
| Contrastive 2 | 9.42 | 0.5018 | 0.5077 | 7.69 | 0.4818 | 0.4951 |

Contrastive 2 system is same with primary system. The difference is that only unlabeled major dataset is used for GL-norm on both development and evaluation experiment.

## 7. CPU EXECUTION TIME

All tasks were performed on 64bit linux with 64G RAM and Intel i7 6700 3.4GHz and GTX1080 for GPU. All CPU times are counted based on one core CPU.

Table 4. CPU Execution time by tasks per 1 utterance corresponding enrollment and test dataset of major language

| Task | Execution time (sec.) | | Memory usage (MB) |
|---|---|---|---|
| | Enrollment | Test | |
| MFCC | 0.61 | 0.39 | 4.9 |
| DNN-MFCC | 0.65 | 0.35 | 5.5 |
| BNF | 24.36 | 8.93 | 6528 |
| VAD | 0.01 | 0.01 | 9.1 |
| GU-i-vector | 5.56 | 5.13 | 3498 |
| DU-i-vector | 43.80 | 24.52 | 13469 |
| SU-i-vector | 8.19 | 7.74 | 8847 |
| BU-i-vector | 5.64 | 5.16 | 4254 |
| PLDA | 0.01 | | 251 |
| IDVC | 0.02 | | 2.5 |
| ILVC | 0.01 | | 2.3 |
| WTLN | 0.01 | | 3.5 |
| SADA | 0.01 | | 34.3 |
| GL-norm | 0.01 | | 1.2 |

Table 5. Total CPU Execution time for 1 trials by systems

| System | Total CPU time for a single trials (sec.) |
|---|---|
| GU-IDVC-ILVC -WTLN-IPLDA-GLnorm | 11.75 |
| DU-IDVC-ILVC -WTLN-IPLDA-GLnorm | 70.38 |
| SU-IDVC-ILVC -WTLN-IPLDA-GLnorm | 16.99 |
| BU-IDVC-ILVC -WTLN-IPLDA-GLnorm | 45.14 |
| DU-SADA-ILVC -WTLN-IPLDA-GLnorm | 70.37 |